\documentclass[twocolumn]{aa} 
\usepackage{graphicx}
\usepackage{txfonts}
\usepackage[T1]{fontenc}
\usepackage{pdflscape}
\usepackage{rotating}
\usepackage{float}
\usepackage{graphicx}
\usepackage{caption}
\usepackage{subcaption}
\begin{document} 

\title{Obscuration in high redshift jetted QSO}

%   \subtitle{I. Overviewing the $\kappa$-mechanism}
   \author{A. Caccianiga\fnmsep\thanks{alessandro.caccianiga@inaf.it}
          \inst{1}
          \and
          L. Ighina
          \inst{1,2,3}
          \and
          A. Moretti
          \inst{1}
          \and
          R. Brivio
          \inst{1,2}
          \and
          S. Belladitta
          \inst{4,5}
          \and
          D. Dallacasa
          \inst{6,7}
          \and
          C. Spingola
          \inst{7}
          \and
          M.J. March\~a
          \inst{8}
          \and
          S. Ant\'on
          \inst{9}
          }

   \institute{INAF - Osservatorio Astronomico di Brera, via Brera 28,  I-20121 Milan, Italy
              %\email{}
         \and
             DiSAT, Universit\`a degli Studi dell’Insubria, via Valleggio 11, 22100 Como, Italy
             %\email{c.ptolemy@hipparch.uheaven.space}
             %\thanks{The university of heaven temporarily does not
             %        accept e-mails}
         \and
         International Centre for Radio Astronomy Research, Curtin University, 1 Turner Avenue, Bentley, WA, 6102, Australia
         \and
         Max Planck Institut für Astronomie, Königstuhl 17, 69117 Heidelberg, Germany
         \and
         INAF – Osservatorio di Astrofisica e Scienza dello Spazio, Via Gobetti 93/3, 40129 Bologna, Italy
         \and
         Dipartimento di Astronomia, Universit\`a di Bologna, via Ranzani 1, 40127, Bologna, Italy
         \and
         INAF - Istituto di Radioastronomia, Via Gobetti 101, I-40129 Bologna, Italy
         \and
         Dept. of Physics\&Astronomy and Dept. of Computer Science,  University College
London, Gower Street, London, WC1E 6BT, UK
         \and
         Centro de F\'isica da UC, Departamento de Física, Universidade de Coimbra, 3004-516 Coimbra, Portugal
             }

   \date{Received XX; accepted yy}

% \abstract{}{}{}{}{} 
% 5 {} token are mandatory
 
  \abstract
   {Obscuration in high-redshift quasi-stellar objects (QSO) has a profound impact on our understanding of the evolution of supermassive black holes across the cosmic time. An accurate quantification of its relevance is therefore mandatory.}
   {We present a study aimed at evaluating the importance of obscuration in high redshift jetted QSO, i.e. those active nuclei characterized by the presence of powerful relativistic jets.}
   {We compare the observed number of radio detected QSO at different radio flux density limits with the value predicted by the beaming model on the basis of the number of oriented sources (blazars). Any significant deficit of radio-detected QSO compared to the predictions can be caused by the presence of obscuration along large angles from the jet direction. We apply this method to two sizable samples characterized by the same optical limit (mag=21) but significantly different radio density limits (30~mJy and 1~mJy respectively) and containing a total of 87 independent radio-loud 4$\leq$z$\leq$6.8 QSO, 31 of which classified as blazars.}
   {We find a general good agreement between the numbers predicted by the model and those actually observed with only a marginal discrepancy at $\sim$0.5~mJy that could be caused by the lack of completeness of the sample. We conclude that we have no evidence of obscuration within angles 10-20$^\circ$ from the relativistic jet direction. We also show how the on-going deep wide-angle radio surveys will be instrumental to test the presence of obscuration at much larger angles, up to 30-35$^\circ$. We finally suggest that, depending on the actual fraction of obscured QSO, relativistic jets could be much more common at high redshifts compared to what is usually observed in the local Universe.}
   {}


   \maketitle
%
%-------------------------------------------------------------------

\section{Introduction}
The study of the very first phase of the supermassive black hole (SMBH) growth represents a critical step for  understanding the galaxy early evolution (e.g. \citealt{Heckman2014,Merloni2015}). To this end, a reliable and complete census of the accreting SMBH, i.e. Active Galactic Nuclei (AGN), at high redshifts is mandatory (e.g. \citealt{Banados2016, Pacucci2021a, Fan2022, Banados2022}). For observational reasons, the study of high redshift accreting SMBHs is mostly limited to the part of the population residing in unabsorbed objects, the so-called type1 AGN, while SMBH embedded in absorbed systems (type~2 AGN)  typically elude observations. Up to now, these elusive objects were nearly beyond the capabilities of the existing telescopes except for few possible examples (e.g. \citealt{Endsley2022, Drouart2020}), although their detection is now becoming possible thanks to the James Webb Space Telescope (e.g. \citealt{Yang2023a}). For these reasons, most of the studies of the SMBH formation and evolution are still based on high redshift unobscured AGN and, therefore, they are potentially biased. Assuming for the high-z AGN the same obscured fraction measured locally is too simplistic, since there are several pieces of evidence showing that the fraction of obscured sources significantly increases at high-z (see e.g. \citealt{Zeimann2011,Moretti2012, Merloni2014,Aird2015,Gilli2022,Vijarnwannaluk2022,Yang2023a}). 

The presence of a relativistic jet in AGN greatly helps to quantify the importance of obscuration. 
When an AGN of this type is observed close to the jet direction, its emission is significantly boosted due to relativistic beaming and the oriented nature of the source can be easily recognized. Such oriented sources are called blazars (see \citealt{Urry1995} for a review). 
Notably, the particular orientation of blazars makes obscuration marginal, since the jet is expected to clear out the path along the line-of-sight. From a statistically complete sample of blazars, selected at a given radio limit, it is then possible to study the impact of the circum-nuclear obscuring medium since the detection of a blazar with a given radio flux density implies the existence of a well-defined number of misaligned sources at lower flux densities ($\sim$2$\Gamma^2$, where $\Gamma$ is the bulk Lorentz factor of the jet): A significant deficit of the observed misaligned sources, compared to the predictions, then implies the existence, at a certain critical angle, of an obscuration structure. Using this approach, \citet{Ghisellini2016} have pointed out the possible presence of an almost 4$\pi$ obscuring structure around the most luminous jetted AGN, although this result was quite uncertain being based on a small sample of blazars.

It is important to note that, from the optical point of view,  the class of blazars includes two different types of sources: 1) BL Lac objects, that are characterized by featureless optical spectra, and 2) flat-spectrum radio quasars (FSRQ), whose optical properties are similar to those observed in radio-quiet quasi-stellar objects (QSO), being usually dominated, at these wavelengths, by the accretion disk emission. The two types of blazars are likely associated to two different classes of jetted AGN, characterized by different physical properties, probably connected to the accretion rate on the central SMBH (e.g. \citealt{Tadhunter2016}). Here, we only consider the class of FSRQ since BL Lac objects are usually not found at high redshifts ($\geq$4) although the possible discovery of a BL Lac at z$\sim$6.57 was recently claimed (\citealt{Koptelova2022a}). Throughout the paper, we will use the term ``QSO" to indicate an AGN with strong emission lines (EW$>$5\AA) and/or a clear evidence of the accretion disk emission in the optical spectrum, regardless of its luminosity and/or the presence of a jet.

In this work, we use two sizable flux-limited samples of high redshift blazars to investigate the issue of obscuration in jetted QSO in the early Universe by adopting a statistical approach similar to that described in \citet{Ghisellini2016}. The first one is the CLASS (Cosmic Lens All Sky Survey) high-z QSO sample (\citealt{Caccianiga2019}; \citealt{Ighina2019a}) which is characterized by a relatively high radio flux density limit (30~mJy) and an optical limit of mag=21, while the second one, based on data from the Faint Images of the Radio Sky at Twenty-Centimeters (FIRST) survey, has a much fainter radio flux density limit (0.5 mJy) but the same optical limit of CLASS. 
The combination of these two samples greatly increases the sensitivity to absorption effects and, at the same time, improves the statistics, making our study more accurate.

In Section~2 we discuss the method used to derive the expected number of jetted QSO from the observed number of oriented sources (blazars) in a flux-limited sample. We then apply this method to the CLASS sample (Section~3) and to the FIRST sample (Section~4).  In Section~5 we discuss the expected improvements offered by incoming new surveys based on SKA precursors/pathfinders. Finally, in Section~6 we summarize our conclusions.

In the paper we use the mag$_{drop}$ defined as the magnitude, in AB system and corrected for Galactic extinction, in the reddest filter of the dropout used to select the sources. This means the {\it r}-filter for 4$\leq$z$<$4.5, the {\it i}-filter for 4.5$\leq$z$<$5.4, the {\it z}-filter for 5.4$\leq$z$<$6.3 and the {\it y}-filter for z$\geq$6.3 objects. By construction, mag$_{drop}$ corresponds to a quite narrow range of rest-frame wavelengths, between $\sim$1250\AA\ and 1450\AA.

Throughout the paper we assume a flat $\Lambda$CDM cosmology with H$_0$=71 km 
s$^{-1}$ Mpc$^{-1}$, $\Omega_{\Lambda}$=0.7 and $\Omega_{M}$=0.3.  Spectral 
indices are given assuming S$_{\nu}\propto\nu^{-\alpha}$.

\section{Blazars versus misaligned sources: a test for obscuration}
In the simplest version of the Unified Model (e.g. \citealt{Antonucci1993}; \citealt{Urry1995}; \citealt{Netzer2015}), QSO are axis-symmetric sources that may appear either absorbed (type~2) or unabsorbed (type~1) depending on whether the line-of-sight intercepts an obscuring structure called dusty torus. 
In this picture, the aperture of the dusty torus determines the observed fraction of absorbed QSO in a given sample. 
Knowing the torus aperture is therefore fundamental in order to account for the (unobserved) obscured population and to obtain a reliable census of the entire population.

The presence of relativistic jets in a fraction of QSO greatly helps to infer the source orientation and, therefore, to assess the existence of an obscuring structure at a certain angle from the jet direction. In particular, the relativistic beaming makes oriented sources orders of magnitudes brighter than misaligned objects and, therefore, easily recognizable thanks to their  ``blazars properties", like a core-dominated, flat spectrum radio emission, variability and a strong X-ray emission with a flat photon index. Since beaming effects are maximized for angles within $1/\Gamma$ from the jet direction ($\Gamma$ is the bulk velocity of the jet), a common assumption is that blazars are observed within this angle, although this statement should be considered valid in a statistical sense. We thus call $\Theta_{b}=1/\Gamma$ the ``blazar angle". This means that for each observed blazar there should be $4\pi/\Omega_b\sim2\Gamma^2$ sources with the same intrinsic properties and with a jet randomly oriented, where $\Omega_b$ is the solid angle corresponding to $\Theta_b$\footnote{The solid angle corresponding to $\Theta_b$ is 2$\pi$(1-cos$\Theta_b$). Since jets are emitted along two opposite directions, $\Omega_b$ is 2 times this value i.e. $\Omega_b=4\pi(1-cos\Theta_b)\sim2\pi/\Gamma^2$ (valid for small angles).}. Therefore, given the number of blazars observed in a certain volume of Universe, it is possible to infer the total number of sources with the same properties within the same volume. The presence of obscuration at a certain viewing angle, reduces the observed number of misaligned radio-emitting QSO, if we are selecting only unobscured, type~1, sources. The comparison between the predicted and the observed number of jetted type~1 QSO can thus yield direct information on the impact of obscuration in the sample.

Real samples, however, are typically flux-limited and the situation is more complicated since beaming greatly boosts the fluxes of blazars, thus favouring their inclusion in the sample.
For this reason, the non-blazar/blazar relative ratio observed in a flux limited survey is, in general, significantly different from 2$\Gamma^2$ and depends on several factors, like the survey limit(s) and the shape of the luminosity function (e.g. see discussion in \citealt{Lister2019}).

 \citet{Ghisellini2016} developed a simple method to predict the expected number of misaligned sources in a flux-density limited sample on the basis of the observed fluxes of blazars selected in the same sample.
 In summary, given a sample with a flux density 
limit S$_{lim}$ and containing N  blazars in a certain redshift interval, we expect to find, at the same
flux limit and within the same redshift interval, a number ($N_{tot}$) of blazars plus non-blazars (i.e. with a misaligned jet) given by\footnote{We note that in \citet{Ghisellini2016} this number is wrongly given as the ratio between N$_{tot}$ and the number of blazars (their eq.~8).}:

\begin{equation}
N_{tot} \sim \sum_{i=1}^N{[2(\frac{S_i}{S_{lim}})^{1/p} -1]}
\end{equation}

where the sum is done on all the blazars selected in the survey and $S_i$ are their flux densities. The parameter ``p'' appears in the beaming model and depends on the jet: for instance, if $\alpha$ is the radio spectral index of the emitting source, we expect p=3+$\alpha$ for a moving, isotropic source and p=2+$\alpha$ for a continuous jet (see Appendix B of \citealt{Urry1995} review paper for more details). It is worth noting that equation~1 does not depend on the value of $\Gamma$, the only free parameter being $p$. This greatly reduces the uncertainties of the method.

We tested the validity of equation~1 via Monte-Carlo simulations and found that it systematically over-predicts the number of sources by a factor $\sim$1.4 (see Appendix~A).  This is likely related to the fact that in the \citet{Ghisellini2016} work all blazars are assumed to be observed at an angle exactly equal to $1/\Gamma$ rather than $\leq1/\Gamma$. If we take this effect into account, and assume that blazars are observed at different angles within $1/\Gamma$, equation~1 becomes (see Appendix~A for its derivation):

\begin{equation}
N_{tot} \sim \sum_{i=1}^N {[1.44(\frac{S_i}{S_{lim}})^{1/p} -1]}
\end{equation}

We will use this formula throughout the paper.

Strictly speaking, the value of N$_{tot}$ provided by equation~2 should be considered 
as a lower limit on the expected number of jetted high-z QSO since it is computed only on the basis
of the jet luminosity, neglecting the un-beamed, extended emission of the source (like that from the radio lobes). 
This effect can be particularly relevant when working with samples selected at low frequencies where the extended emission is expected to be more important, while it should progressively become less relevant when moving to higher frequencies. When dealing with high redshift  (z$>$4) sources, where a typical observed frequency of 1.4~GHz corresponds to rest-frame frequencies above 7~GHz, the relevance of the extended components should be marginal. In addition, the extended radio emission is expected to be partially dumped by the interaction between the electrons in the jets and the photons from the Cosmic Microwave Background (CMB, e.g. \citealt{Ghisellini2009b, Paliya2020, Ighina2021c}). In any case,  in the analysis described in the next sections we will always use the peak flux densities, rather than the total integrated flux densities, in order to minimize the possible contribution from any extended emission. 

Having derived and tested a method to estimate the expected number of type~1 jetted QSO in a radio flux-limited sample, we can apply it to real samples to see if this number is consistent with the observations. A deficit of the observed number may suggest the presence of obscuration at a given angle from the jet direction. If the sample is large enough, we can even apply equation~2 at different flux densities. In this way it is also possible to infer the angle at which the obscuration occurs, as it will be shown in the next sections.

We first apply the method to the CLASS sample, characterized by a high radio flux limit compared to the optical one. We then build a much radio deeper sample that contains more misaligned objects, making the comparison between predictions and observations more stringent.  

\section{The CLASS sample}
The first sample considered is the CLASS sample of high-z QSO (\citealt{Caccianiga2019}, \citealt{Ighina2019a}). This is a completely identified flux limited sample of flat-spectrum sources with z$\geq$4 at the radio limit of 30~mJy at 5~GHz and mag$_{drop}\leq$21, covering 13120 sq. degrees of sky at high Galactic latitude ($|b^{II}|>$20$^\circ$). The original complete sample presented in \citet{Caccianiga2019} contained 21 spectroscopically confirmed z$>$4 QSO, out of which 18 were classified as blazars on the basis of their X-ray emission (see \citealt{Ighina2019a} and the Section~4.1 for more details on the classification method). The CLASS sample has been used to obtain an accurate estimate of the space density of high-z blazars (\citealt{Caccianiga2019}; \citealt{Ighina2021c}) and to track the evolution of the most massive SMBH (above 10$^9$ M$_\odot$)
hosted by jetted QSO up to z$\sim$6 (\citealt{Diana2022}).

We now add two sources (J222032.5+002537 at z=4.1960 and HZQJ142048.0+120546 at z=4.0344), both classified as blazars (\citealt{Sbarrato2014a}), that were not originally included because their radio spectrum between 1.4 and 5GHz was steeper than 0.5 and, therefore, above the threshold used to define CLASS ($\alpha_R\leq$0.5)\footnote{These objects are the only known z$>$4 QSO in the sky area covered by CLASS with a flux density at 5~GHz above 30~mJy. In principle, there could be more sources not yet discovered as high-z QSO. However, most of the high-z QSO in the northern sky and with mag$\leq$21 should have been already found, thanks to the Sloan Digital Sky Survey (SDSS) spectral database (see also discussion in Section~4).}. Therefore, the final sample considered here contains 23 sources, 20 classified as blazars and 3 as misaligned objects (see Tab.~\ref{tab:sample}).
At first glance, this dominance of blazars in the CLASS sample could be considered as an evidence of the fact that obscuration plays an important role in hiding misaligned sources. However, as previously discussed, the actual fraction of blazars in a flux limited sample depends on several factors and only using the method described in Section~2 we can establish whether the observed number is consistent or not with the expectations.

If we apply equation~2, where N=20 is the total number of blazars and $S_i$ are their 5~GHz flux densities reported in Tab~\ref{tab:sample} (column~7), we predict 
the existence of 20-24 QSO (using p=3 and 2 respectively) in the same area and with the same radio and magnitude limits, that is 5 misaligned sources at most. This is fully consistent with what we observe.
Therefore, the large fraction of blazars observed in the CLASS survey is not surprising, and it is simply due to the high radio flux limit compared to the optical one, something that favours the selection of sources with high radio-to-optical flux ratio i.e. oriented sources. 

The fact that CLASS is not very sensitive to sources observed at large viewing angles implies that we are only able to test the presence of obscuration at angles slightly above $1/\Gamma$. In order to extend the test at large angles, it is necessary to lower the radio flux limit, while keeping the same optical limit. In this way, we will start selecting more sources with lower radio-to-optical flux ratio values that are likely observed at larger viewing angles. In the next section, we discuss the selection of a sample that is much deeper in the radio than CLASS (by more than a factor 10) but has the same optical limit.

\section{The FIRST sample}
In order to extend the analysis at lower flux densities, we have considered the sample of z$>$4 QSO detected in the FIRST catalogue (\citealt{Becker1995}). The sample was selected by cross-matching FIRST sources with all the objects in the SDSS-DR17 database spectroscopically classified as ''QSO" with z$\geq$4, using the provided spectroscopic search tool\footnote{https://skyserver.sdss.org/dr17/SearchTools/SQS}. We then complemented this search by considering all the known QSO from the literature with a spectroscopically confirmed redshift above 4, using either NASA/IPAC Extragalactic Database\footnote{http://ned.ipac.caltech.edu/} or SIMBAD\footnote{http://simbad.cds.unistra.fr/simbad/}.
To obtain a reliable quantification of the number of z$>$4 QSO we restricted our analysis to a specific sky area (9h$\leq$RA$\leq$16h, 0$^\circ\leq$DEC$\leq$60$^\circ$, $|b|>20^\circ$) where we expect that most of the z$\geq$4 and mag$_{drop}\leq$21 have been actually identified thanks to the SDSS spectroscopy. 
We have verified that in this area, that covers 5215 sq. degrees, all the high-z sources with mag$\leq$21 present in CLASS have an SDSS spectrum in the last data release. For this reason, we are confident that the completeness of the sample is high ($>$90\%), at least up to z$\sim$5.5. This is in agreement with the results from \citet{Schindler2017a} showing that in this sky area and for  relatively faint magnitudes ($>$19) the QSO selection should be highly complete.

The SDSS spectra of all the z$>$4 QSO from DR17 falling in this sky area have been visually inspected in order to confirm the redshift or exclude objects with a wrong z estimate (most of which claimed to be at z$>$5). In addition, we have searched for all the sources in the literature that have been classified as z$\geq$4 QSO (with a spectroscopic observation) in the same sky area.
We have thus obtained a list of 1330 spectroscopically confirmed z$\geq$4 QSO with mag$_{drop}\leq$21, 66 of which are detected in the FIRST catalogue ($S_{1.4GHz}^{peak}>$1 mJy). Sixty-two of these objects have a spectrum from SDSS, while 4 sources have been found from the literature (see Tab.~\ref{tab:sample}). Two of these (J090132.6+161506 at z=5.63, \citealt{Banados2015a} and J112925.3+184624 at z=6.82, \citealt{Banados2021}) are at z>5.5 where the completeness of the SDSS is known to be lower. The other two objects  (J101337.8+351849, \citealt{Gloudemans2022} and J145224.2+335424, \citealt{Stern2000}) have redshift below 5.5. 
Overall, the sources that are not found by the SDSS represent $\sim$6\% of the sample ($\sim$3\% considering the 4$\leq$z$\leq$5.5 range) and this is consistent with the hypothesis that the SDSS spectroscopic sample is highly complete $>$90\%) in this area of sky and for sources in this range of redshift and magnitude.

\begin{figure*}
\centering
\includegraphics[width=13cm]{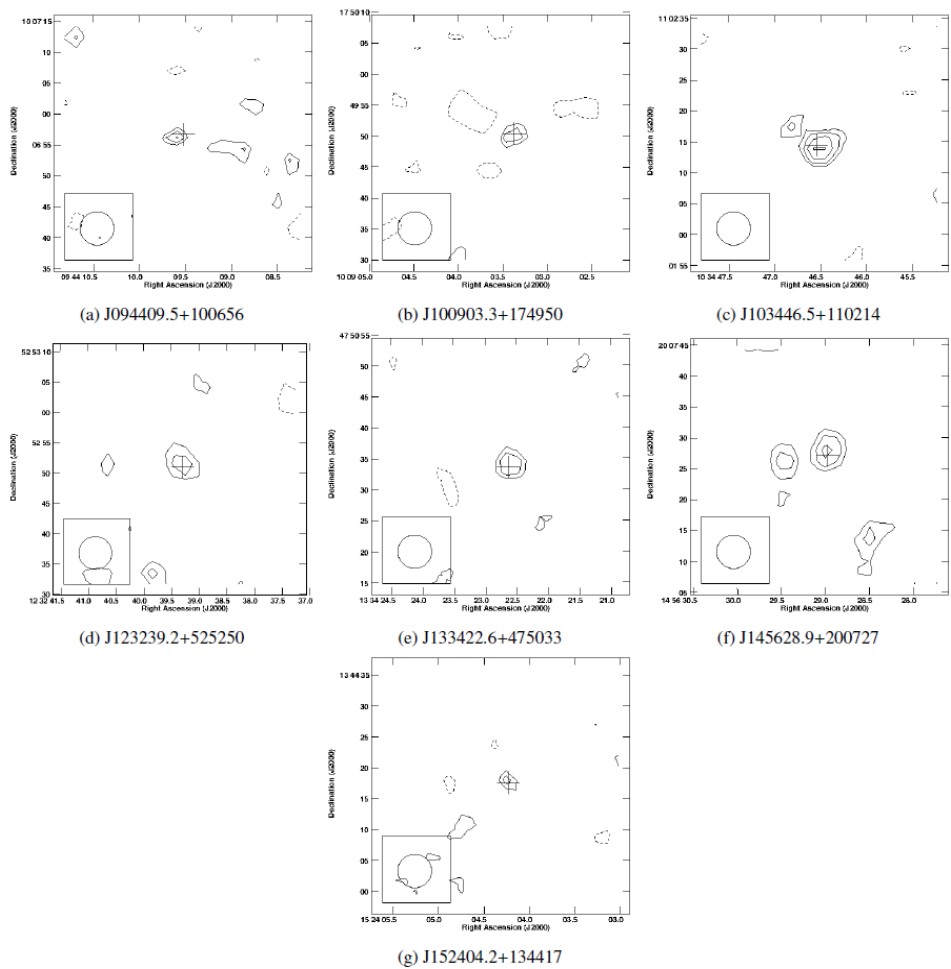}
	\caption{Radio maps at 1.4~GHz from FIRST data of the 7 high-z QSO in the sample with peak flux densities between 0.5 and 0.9 mJy/beam, not included in the FIRST catalogue. Levels are computed as -2, 2, 2.8, 4, 5.5, 8, .. times the rms (0.15 mJy/beam). The beam size is indicated in the bottom-left corner.}
 \label{fig:maps}
\end{figure*}

Since the FIRST catalogue has been produced using a 5$\times$rms threshold, it is possible to extend the sample down to lower flux densities using the FIRST radio data and searching for S$_{1.4GHZ}^{peak}\geq$0.5 mJy/beam (>3$\times$rms) around the optical position of z$>$4 QSO. This lower threshold is reasonable considering that we are "forcing" the photometry towards a limited number of known positions and not carrying out a blind search, as the original FIRST catalogue. 
Using this technique, we have found 7 additional sources with peak surface densities between 0.5 and 0.9~mJy/beam. One of these sources (J133422.6+475033) has been also detected at 140~MHz in the Low Frequency Array (LOFAR) Two-metre Sky Survey (LoTSS, \citealt{Gloudemans2022}). The radio maps of these newly discovered radio detections are reported in Fig.\ref{fig:maps}. Since it is difficult to test the actual completeness of this extension at 0.5~mJy we use the original sample at 1~mJy for most of the computations while we use the extension only to reveal hints of possible trends at lower flux densities.

In total, the final FIRST sample contains 73 z$\geq$4 radio-emitting QSO with a peak flux density at 1.4~GHz $\geq$0.5 mJy/beam and mag$_{drop}\leq$21.
In Tab.\ref{tab:sample} these sources are flagged with the letter "F".
These 73 radio detections correspond to 5\% of the total number of z$\geq$4 type~1 QSO with mag$\leq$21 present in the considered sky area (1330).

Nine sources in the FIRST sample are in common with CLASS (i.e. all the CLASS high-z QSO falling in the sky area covered by FIRST). Therefore, the two samples studied here contain a total of 87 independent z$\geq$4 jetted QSO.

\subsection{Blazar classification}
 In order to apply eq.~2 it is necessary to recognize all blazars present in the FIRST sample. Historically, blazars with emission lines were identified with the class of FSRQ that are defined on the basis of their flat ($\alpha_R<$0.5) radio spectrum. A flat spectrum is usually attributed to a source that is dominated by the self-absorbed emission from the core, i.e. from the relativistically boosted, unresolved part of the jet (e.g. \citealt{Urry1995}). Often, the slope of the radio spectrum is evaluated on the basis of non-simultaneous flux densities measured at two different frequencies. Not only variability may affect the measure of the slope, but also complex spectral shapes, like peaked spectra, can lead to a mis-classification of a source as FSRQ if only two points are used to characterize the spectrum. This is particularly true for high-z jetted QSO that often show gigaherts-peaked spectra, possibly due to their young age (e.g. \citealt{Frey2011, Momjian2021, Zhang2021, Belladitta2023a}). Moreoever, at high redshifts we usually observe at high rest-frame frequencies where the core, that has a flat spectrum, can dominate over the steep-spectrum extended emission even in moderately mis-aligned sources. This can be even more true considering that the extended emission can be dumped at high redshift due to the effect of the CMB, as mentioned in Sect.~2.  
 For all these reasons, while the two-points radio spectrum can be effectively used to efficiently pre-select high-z blazar candidates (e.g. \citealt{Caccianiga2019}), caution should be used in adopting the radio slope to classify a source, in particular when the spectrum is measured with non-simultaneous flux densities. 
 The availability, in the next future, of surveys like GLEAM-X (\citealt{Wayth2018, Hurley-Walker2022}) that will provide simultaneous spectra between 72 and 231 MHz (corresponding to rest-frame frequencies of 400MHz-1.3GHz at z=4.5) for sources down to 5~mJy, will allow a more accurate spectral classification of high-z radio-emitting QSO. 

 Alternatively (or in addition), high-resolution, VLBI observations are often used to constrain the orientation of a high-z QSO. In particular, the detection of a high brightness temperature (T$_b$) above the equipartition limit, is usually considered a robust way to classify a source as blazar (\citealt{Frey2008,Frey2010,  Gabanyi2015, Coppejans2016a, Frey2018, Spingola2020d, Liu2022a, Krezinger2022a}). Unfortunately, for sources of a few mJy (or below) the maximum baseline available on earth is not long enough to put a stringent limit on T$_b$ able to unambiguously classify a source as blazar. Also, the sensitivity could be an issue for very faint sources. Even for brighter objects, the estimate of T$_b$ requires an accurate measure of the source size that is often only a small fraction of the synthesized beam, and this may be a challenging task. Finally, the systematic follow-up with VLBI techniques of large samples of high-z QSO can be particularly time-consuming. 

 A third method to classify a source as blazar is based on the analysis of the Spectral Energy Distribution (SED). A SED showing a strong and "rising" X-ray emission (e.g. with a Photon index below 1.5), well above the one expected for a non-jetted QSO, due to the electrons of the hot-corona, is considered a clear signature of the orientation of the source (\citealt{Ghisellini2010, Sbarrato2012, Sbarrato2013, Ghisellini2019, Sbarrato2021a, Sbarrato2022}). 
 In \citet{Ighina2019a} we have used a similar method, based on the X-ray-to-optical luminosity ratio, parametrized by the $\tilde{\alpha}_{ox}$ index\footnote{According to \citet{Ighina2019a} we define 
$\tilde{\alpha}_{ox} \, = - \, \frac{ log(L_{10keV} \, / \, L_{2500\AA})}{log(\nu_{10keV} \, / \, \nu_{2500\AA})} = -0.3026 \, log(\frac{L_{10keV}}{ L_{2500\AA}})$, where $L_{10keV}$ and $L_{2500\AA}$ are, respectively, the X-ray and UV monochromatic luminosities (per unit of frequency) Blazars are characterized by $\tilde{\alpha}_{ox}\leq$1.355.}, and on the X-ray spectral slope 
to quantify the X-ray dominance in the SED and to easily distinguish blazars from mis-aligned sources. If the X-ray spectral index is not available, or poorly determined, the $\tilde{\alpha}_{ox}$ parameter alone can be effectively used.

 The different methods described above do not always agree in classifying the sources (e.g. see discussion in  \citealt{Krezinger2022a}) and it is likely that the most reliable classification for a single object will be achievable only by combining all the pieces of information available from radio to X-rays. For large samples, however, using a single method like the one based on X-ray data can represent a reasonable compromise to easily obtain a uniform classification that is sufficiently reliable, at least from a statistical point of view. For this reason, we adopt the X-ray method to classify the sources in the FIRST sample. This is the same method adopted for the CLASS sample as explained in the previous section and in \citet{Ighina2019a}. In Appendix~B we discuss the reliability of the adopted classifications by comparing them with those based on SED modelling and VLBI data from the literature. In Sect.~4.2 we will evaluate the impact of possible mis-classifications on the final results.
 
 For radio flux densities above 30~mJy nearly all sources are included in the CLASS sample of high-z QSO and, therefore, we can use the classifications obtained in \citet{Ighina2019a}. For lower flux densities, we expect a smaller number of blazars since the average radio-to-optical flux ratio decreases progressively when lowering the radio flux density limit and keeping the same optical limit. 
In order to apply the same criteria used in \citet{Ighina2019a} to classify the sources in the FIRST sample, we exploited all the available sources catalogues of the major existing X-ray telescopes ({\it XMM-Newton, Chandra, Swift-XRT, NuSTAR}). This search has provided X-ray data for 24 objects. 
In particular, for 18 sources we have fluxes from the  Chandra Source Catalog Release 2.0 (CSC 2.0, \citealt{Evans2010}) and for 6 additional objects we have obtained an X-ray flux from Swift XRT Point Source Catalogue (2SXPS, \citealt{Evans2020}). The Chandra 0.5-7.0~keV fluxes, computed using a power-law with a fixed photon index of 2.0, and the Swift-XRT 0.3-10~keV fluxes, computed assuming a power-law with a fixed photon-index of 1.7, all corrected for the Galactic absorption, are then converted into the rest frame monochromatic 10~keV flux assuming the same photon indices mentioned above. The $\tilde{\alpha}_{ox}$ is finally derived by combining the computed X-ray flux at 10~keV and the monochromatic flux at 2500\AA\ (rest frame) estimated from the z magnitudine and using the optical spectral index computed between WISE W1 and z magnitudes, if available, or assuming $\alpha_o$=0.44 (\citealt{VandenBerk2001}).

For one additional object (J091316.5+591921), not present in the two catalogues mentioned above, we have derived the value of $\tilde{\alpha}_{ox}$ from the $\alpha_{ox}$ (defined between 2500\AA\ and 2~keV) published in \citet{Wu2013} and using the conversion formula reported in \citet{Ighina2019a}.  For one more object (J090132.6+161506) we have analyzed a 13ks public Chandra observation (PI: Garmire) and computed the X-ray flux. Finally, for 2 additional sources, not included in the CSC2.0 and 2XSPS catalogues but for which there is a Chandra or Swift-XRT observation, we have derived an upper-limit on the X-ray flux and, consequently, a lower limit on $\tilde{\alpha}_{ox}$.

In total, we have obtained X-ray data for 28 objects that represent 38\% of the sample (see~\ref{tab:sample}). 
The fraction of sources without a classification is therefore quite high (62\%).
It should be noted, however, that not all the sources without X-ray data are reasonable blazar candidates. Blazars are typically characterized by high radio-to-optical luminosity ratios.
We quantify the radio-to-optical relative strength of a source adopting the commonly used radio-loudness parameter (R) defined as (\citealt{Kellermann1989b}):  R=$\frac{S_{R}}{S_{O}}$, where $S_{R}$ and $S_{O}$ are the k-corrected flux densities at the rest-frame frequency 5~GHz and 6.81$\times$10$^{14}$Hz respectively (corresponding to a wavelength of 4400\AA).
Typically, blazars have high 
radio-loudness values ($\gtrapprox$100 e.g. \citealt{Sbarrato2014a}).  Indeed, considering all the z$\geq$4 QSO currently detected at radio wavelengths and with a classification based on X-ray data, we notice that the large majority (93\%) of the sources with R$\geq$1000 are classified as blazars on the basis of the X-rays. Vice versa, only $\sim$15\% of the sources with R$<$80 have values of $\tilde{\alpha}_{ox}$ consistent with those of blazars (most of which close to the 1.355 threshold).  
In the intermediate region (80$\leq$R$<$1000) we have a mix of possible classifications. This means that, if X-ray data are missing, only sources with a radio-loudness between 80 and 1000 really need to be observed while the remaining sources can be classified as blazars or non blazars only on the basis of their radio-loudness (at least with a $\sim$90\% confidence level). Out of the 45 objects in the FIRST sample without X-ray data, only 10 have 80$\leq$R<$1000$ while the remaining have R$<$80. 
We have a running project aimed at observing these 10 sources with Swift-XRT. More X-ray observations will be available when data from the ongoing {\it eROSITA} all-sky survey (\citealt{Merloni2012a}) will be released. For the time being, we will estimate the impact of the sources with missing classification to the final results (Sect.~4.2).

Table~\ref{tab:sample} summarizes the properties of all the sources in the sample including, when available, a classification as blazar/non-blazar based on X-ray data. In particular, 20 sources have been classified as blazars in the FIRST sample. Since nine of these sources are in common with the CLASS sample, Table~\ref{tab:sample} contains a total of 31 high-z QSO classified as blazars.

\subsection{Results}
%-------------------
   \begin{figure}
   \centering
    \includegraphics[width=9cm, angle=0]{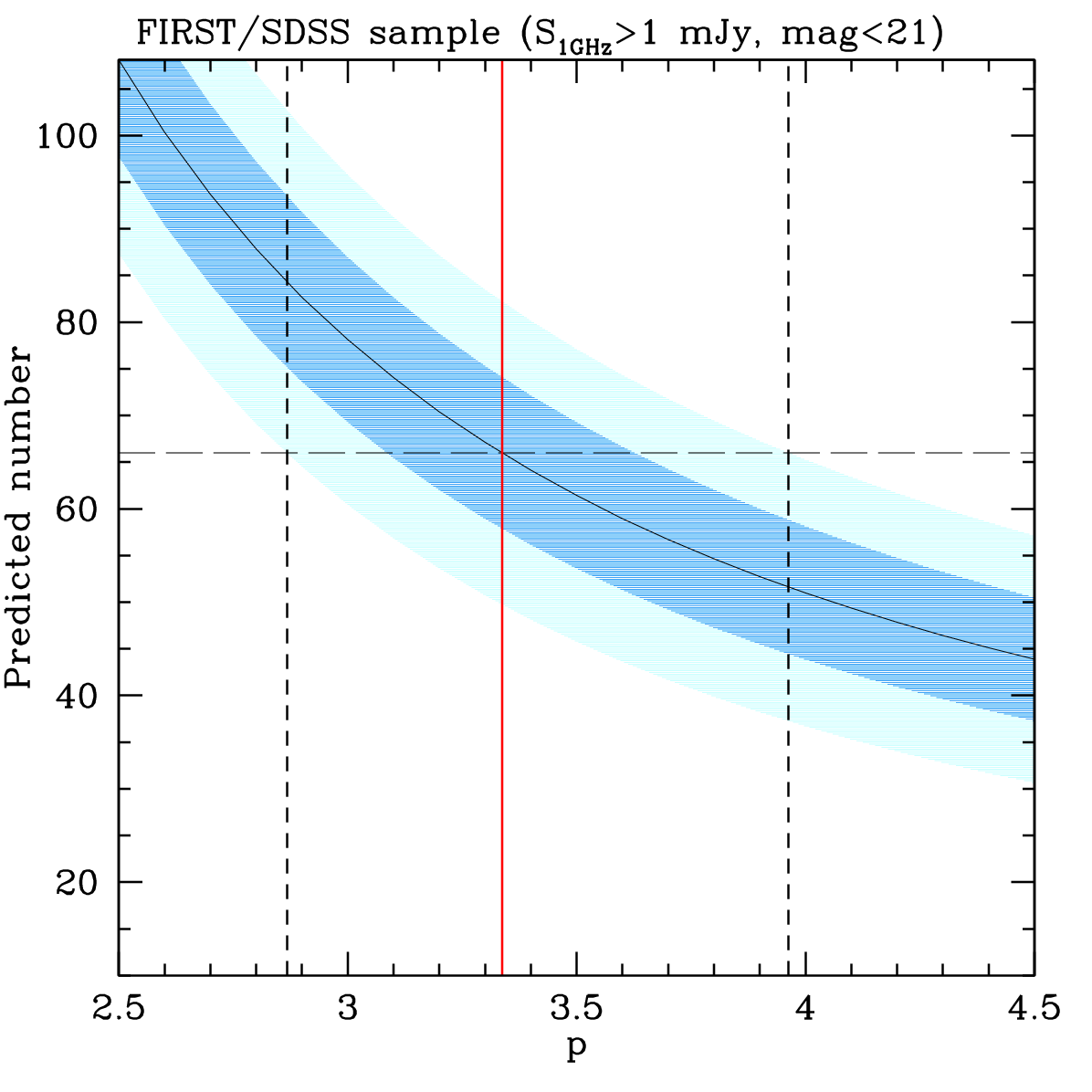}
    \caption{Predicted number of z$\geq$4 QSO (both blazars and non-blazars) with flux density at 1.4~GHz $\geq$1mJy and mag$_{drop}\leq$21 in the sky area covered by the FIRST sample, based on the observed number of blazars in the same area. The predictions are shown as a function of the parameter "p" while the shaded area shows the Poissonian uncertainty on this number (dark blue=1$\sigma$, light blue=2$\sigma$). The horizontal dashed line indicates the observed number of sources, while the vertical lines report the best value of p (red solid line) together with the 2$\sigma$ confidence interval (black short-dashed lines).}
              \label{first_number_torus}
    \end{figure}
%----------------------------------------------
We can now apply equation~2 to the blazars selected in the FIRST sample to predict the expected number of radio-emitting high-z QSO. 
In Fig.~\ref{first_number_torus} we show the expected number of sources with a peak flux density above 1~mJy as a function of the parameter "p", which is the only free parameter in equation~2. We obtain a number consistent with the one observed for a reasonable value of p ($\sim$3.3) with a [2.9-4.0] 2$\sigma$ interval. Considering that the average spectral index of the sample is $\sim$0.1-0.4, depending on the frequencies, the value of p=3.3 is consistent with the scenario of a moving, isotropic source (p=3+$\alpha$, \citealt{Urry1995}). 
We note that the inclusion of the unclassified sources as blazars in the analysis would increase the derived value of p. If we consider as blazars all the objects without X-ray data but with a radio-loudness higher than 80 (see discussion in the previous section) we obtain a value of p of 3.8. This is an extreme case, since we do not expect that all these objects are blazars. Also, the possible misclassifications can have an impact on the best-fit value. If we exclude all blazars whose classification is not supported by VLBI data (see Appendix~B) we obtain a slightly lower value of p (3.0). Realistically, the actual value of p should be included between 3 and 3.8. Since all these values of p are reasonable, we conclude that no significant departures from the beaming model expectation are observed, even considering the uncertainties related to the delicate issue of the blazar classification.

%______________________________________________ PREDICTIONS
   \begin{figure}
   \centering
    \includegraphics[width=9cm, angle=0]{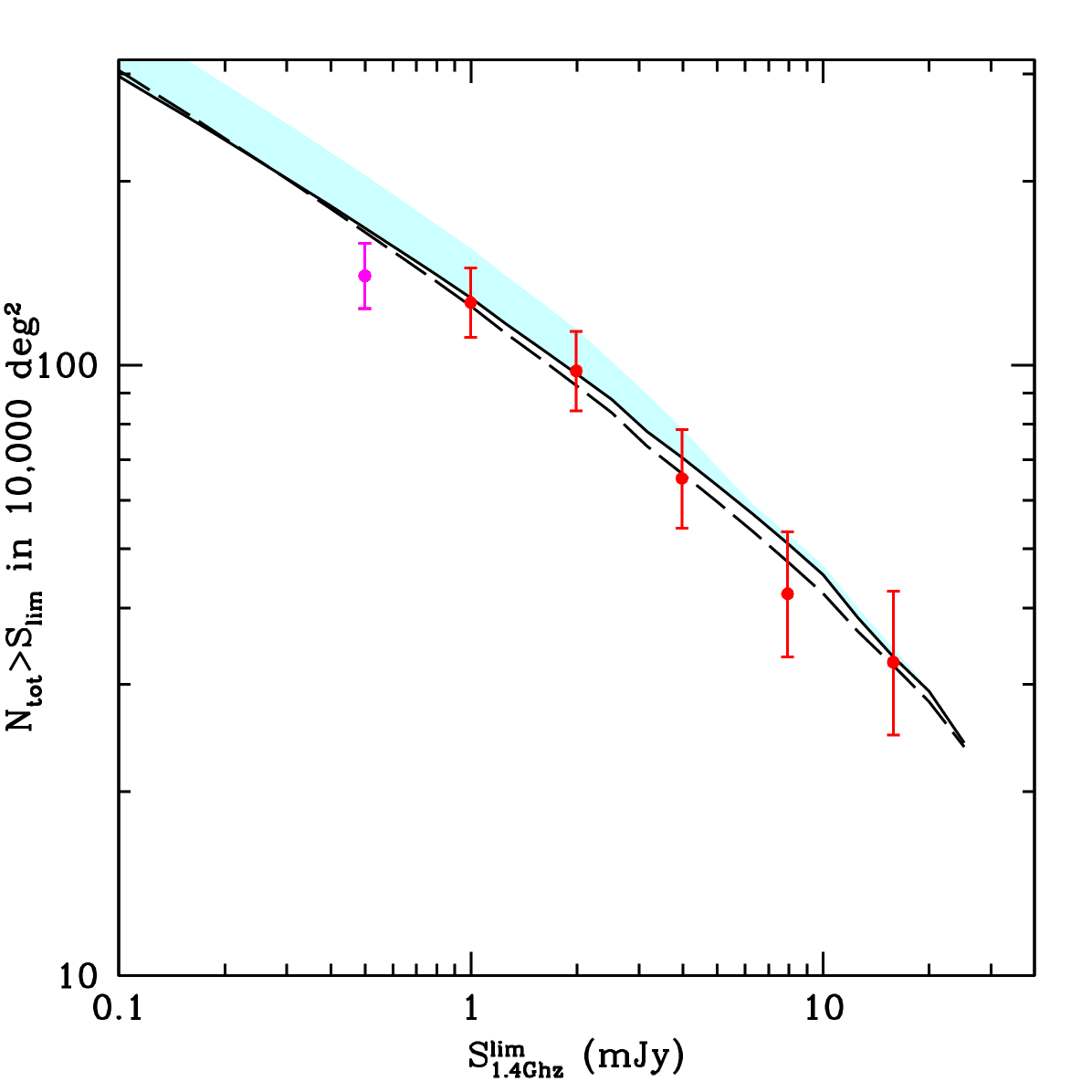}
   \caption{Predicted number, on a sky area of 10 000 sq. degrees, of z$\geq$4 jetted type1 QSO (blazars+mis-aligned sources) with mag$_{drop}\leq$21 for different radio flux density limits, based on the high-z blazars currently detected at 1.4~GHz in the FIRST sample (mostly at flux densities above 10~mJy) described in the text. The predictions (black solid line) assume the best value of p (3.3). The light-blue shaded area indicates the possible impact of the sources with missing classification under the (extreme) assumption that all the sources missing X-ray data and with a radio-loudness above 80 are blazars.
   The dashed line, instead, shows the impact of the mis-classifications: the line is computed by excluding all blazars that are not confirmed by VLBI observation (see Appendix~B). In this case, we assume p=3.0.
   Red points represent the observed sources from the FIRST catalogue while the magenta point indicates the extension down to 0.5~mJy/beam based on FIRST maps.
}              \label{fig_first_gamma10_unc}
    \end{figure}

%______________________________________________ PREDICTIONS
   \begin{figure}
   \centering
    \includegraphics[width=9cm, angle=0]{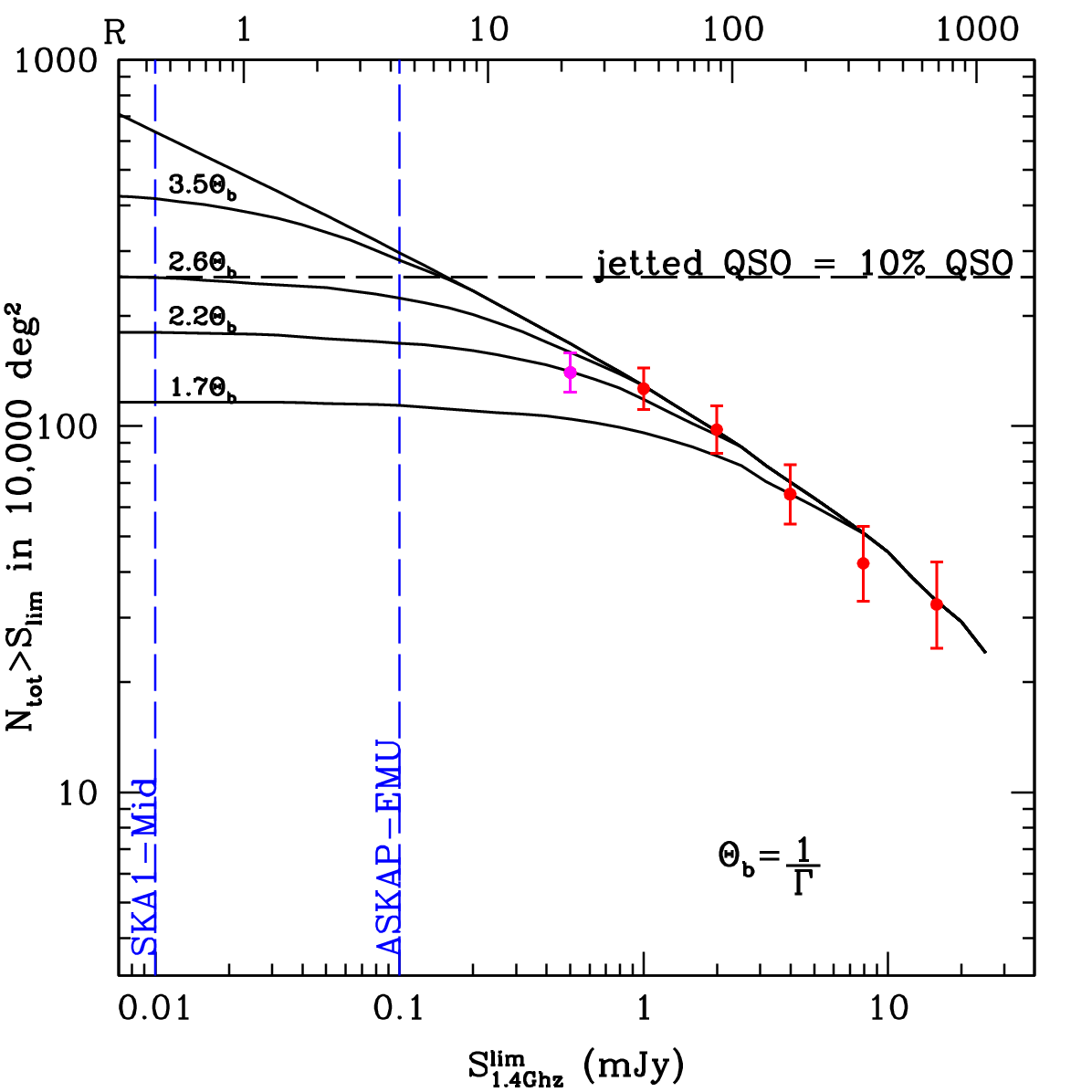}
   \caption{As figure~\ref{fig_first_gamma10_unc} but showing the impact of a circumnuclear obscuring torus with different aperture angles expressed in units of beaming angle $\Theta_b=1/\Gamma$ (the uppermost line represents the case of no-absorption). The upper scale of the figure shows the representative values of radio-loudness (R) corresponding to the flux density limit in the X-axis and using a magnitude of 21 and assuming an average redshift of 4.5. The depth (5$\times$rms) of some ongoing/planned surveys is also indicated. Points as in previous figure. The horizontal dashed line indicates the expected total number of jetted QSO in 10,000 sq. degrees if they represent 10\% of the total QSO population.
   } \label{fig_first}
    \end{figure}
%______________________________________________ PREDICTIONS
Another way of applying eq~2 is to split the sample at different flux limits and predict the expected number of jetted sources at each flux density. In this way, we do not just compare the total predicted number of sources with the observed one, but we also test the distribution of the numbers as a function of the flux limit. In addition, if we are confident that no - or very few - blazars with mag$_{drop}\leq$21 are present at flux densities below the FIRST limit, we can even extrapolate this computation to very flux densities since the presence of blazars at higher flux densities implies the existence of well-defined numbers of misaligned objects at much lower flux densities.  
This is shown in figure~\ref{fig_first_gamma10_unc} where we have assumed the best-fit value p=3.3.
In this figure, we have quantified the impact of the sources without a classification as blazar or non-blazar by assuming the extreme scenario in which all the unclassified sources with a radio-loudness above 80 are blazars: this corresponds to the upper part of the light-blue area reported in the figure. 
Realistically, the correct curve should fall within the blue shaded area, probably closer to the lowest line (corresponding to the case where no blazars are present among the unclassified objects). We have also evaluated the impact of the blazar classifications that are not confirmed by VLBI data (see Appendix~B) by excluding them from the analysis (dashed line) and assuming p=3.0. As clear from the figure, 
the impact of the possible mis-classifications is marginal.

Overall, the shape of the distribution seems in good agreement with the predictions, considering the statistical errors, although the last point, at 0.5~mJy, suggests a possible flattening of the curve. This hint of flattening, however, is statistically marginal, and it could be related to some incompleteness of our extension of the FIRST catalogue below 1~mJy. More data are required in order to confirm this possible trend. 

In case of a significant deficit of sources observed at low flux densities, compared to the predictions, it will be possible to infer the presence of an obscuring structure and to quantify its angular aperture. For instance, according to the AGN Unified Model we expect that above a certain viewing angle the line-of-sight will intercept the molecular torus that will obscure the innermost QSO emission like that from the Broad Line Region (BLR) and the accretion disk. The consequence will be a progressive paucity of observed type~1 QSO (the class of sources that we are considering here) as we move towards lower radio flux densities, where sources are observed - on average - at larger angles.

If we take into account this effect, we expect a flattening of the number of jetted (type~1) QSO below a critical flux density that depends on the torus aperture. 
We note that, while the number of jetted QSO predicted according to the equation~2 does not depend on the Lorentz factor, the flux density limit at which we expect the flattening does. Indeed, for a given torus critical aperture, the corresponding observed flux density depends on the jet bulk velocity: for larger (lower) values of $\Gamma$ the break on the number density will be observed at lower (larger) flux densities. 
In Fig.~\ref{fig_first} we report the expected number of jetted type~1 QSO at z$\geq$4 and mag$_{drop}\leq$21 in a sky area of 10000 sq. degrees at different flux density limits considering different torus apertures expressed in units of ``blazar" angle $\Theta_b$=1/$\Gamma$. 

Assuming that there are no blazars among the unclassified sources, the data points currently do not suggest the presence of a significant level of obscuration at angles lower than $\sim$2 times the critical angle $\Theta_b$ i.e. below 10-20$^\circ$, for $\Gamma$=10-5 respectively. The detection of some new blazars among the unclassified source may increase the discrepancy between the point at 0.5~mJy and the predictions, although, as explained above, this point may suffer from some incompleteness. 
Deeper radio surveys are necessary to better quantify the number of sources at flux densities below 1~mJy and to extend this test to even lower fluxes, corresponding to larger viewing angles (see Sect.~5).

\subsection{Fraction of jetted QSO and absorption}
Since we built the FIRST sample on a sky area where the total number of z$\geq$4 QSO (both radio-quiet and radio-loud) with mag$_{drop}\leq$21 is known, mostly thanks to the SDSS spectroscopic archive, it is possible to estimate the fraction of jetted QSO at these redshifts.

The LogN-LogS presented in Fig.~\ref{fig_first} is expected to converge, at very low flux densities, to the total number of jetted QSO i.e. 2$\Gamma^2\times$N$_{blazar}$ (see discussion in Sect.~2). Since we are dealing only with un-obscured sources, the actual number is:
\begin{center}
2$\Gamma^2\times$N$_{blazar}\times (1-f_{obsc})$
\end{center}
where $f_{obsc}$ is the fraction of obscured sources (N$_{obsc}/N_{tot}$). 
Therefore, the existence of 20 blazars in this area and with mag$_{drop}\leq$21 implies the existence of $\sim$(1000-4000)$\times$(1-f$_{obsc}$) jetted type~1 QSO with the same optical limit, using $\Gamma$=5-10 respectively. 
If we assume that the fraction of obscured (type~2) jetted QSO is similar to that proposed for the total QSO population ($\sim$70\%, \citealt{Vito2018}) we expect, in this sky area,~300-1200 jetted type~1 QSO, most of which not detectable in the existing radio surveys due to their very low radio flux. 
In the same area of sky we have counted 1330 
z$\geq$4 QSO, independently to the radio detection (see Sect.~4).
Therefore, depending on the actual value of $\Gamma$, a significant fraction (23-90\% for $\Gamma$=5 and 10 respectively) of QSO at z$\geq$4 could have a powerful relativistic jet currently not detected because  strongly de-beamed. This result is consistent with what has been found by \citet{Diana2022} using blazars in the CLASS survey (see their figure~5) but it is in apparent disagreement (in particular for high Lorentz factors) with the results presented by \citet{Liu2020a} who found a radio-loud fraction at z$\sim$6 close to 10\% (9.4$\pm$5.7\%). We note, however, that a direct comparison of this number with the fraction of jetted QSO derived above is difficult since our estimate of the fraction of jetted QSO is not based on the value of radio-loudness, as in \citet{Liu2020a}, who used R$>$10 to define a QSO as radio-loud. In our estimate, instead, the misaligned jetted QSO that we infer from the observed number of blazars do not necessarily have R$>$10 since their core emission is expected to be significantly de-beamed at large viewing angles (see Sect.~5.2 for a discussion). And, indeed, seven objects in the FIRST sample have a radio-loudness between 1 and 10. In this sense, the definition of radio-loud QSO has a somewhat ambiguous meaning, in particular when the radio flux is dominated by the core emission which is strongly dependent on orientation (see also the discussion in \citealt{Sbarrato2021a}). 

Alternatively, if we assume that the fraction of jetted QSO is similar to what is found in \citet{Liu2020a} ($\sim$10\%) the fraction of obscured sources must be much higher than 70\%, of the order of 87\%-97\%. Therefore, assessing the importance of obscuration in high-z jetted QSO is intimately related to the problem of how ubiquitous powerful relativistic jets were in the primordial Universe. 

In order to distinguish between these scenarios (high fraction of jetted QSO or high fraction of obscured QSO) we need to reach low flux densities.  In figure~\ref{fig_first} we show the case where jetted QSO represent 10\% of the total population (dashed horizontal line). This is the value to which the logN-logS should converge at low radio flux densities if the fraction of jetted QSO is really 10\%. Reaching flux densities of $\sim$0.1~mJy would already allow us to confirm or reject this possibility. This is a feasible task that can be accomplished in the next few years, as discussed in the next section.

\section{Predictions at sub-mJy flux densities}

Next  years will be particularly promising as far as the radio surveys are concerned. Square-Kilometre Array (SKA) precursors and pathfinders and, subsequently, SKA Observatory (SKAO) itself, are carrying out, or planning, several continuum surveys that will significantly improve the existing ones. From figure~\ref{fig_first} it is clear that we need surveys covering a significant portion of the sky (at least a few thousands of sq. degrees, in order to find a sizable number of objects) reaching flux density limits below 100~$\mu$Jy/beam where we may expect to see a flattening of the high-z RL QSO number counts. A critical point, when going to such a low flux density level, is that the core emission may not be dominant any more because of the strong de-beaming expected at large viewing angles. The presence of a significant fraction of extended emission\footnote{It is however possible that extended emission in the early Universe was strongly dumped due to the interaction with the CMB, as mentioned in Sect.~2.} may cause the inclusion in the sample of more sources than expected on the basis of the sole core emission, thus affecting the analysis. Using high frequencies may help to limit but not completely eliminate this problem. Therefore, a good spatial resolution, of less than a few arcsec, will also be very important in order to resolve the source and distinguish the core from the extended emission.

At $\sim$1~GHz frequency, a promising survey is the "Evolutionary Map of the Universe" (EMU) that is ongoing at the Australian SKA Pathfinder (ASKAP) radio telescope (\citealt{Norris2011b,Norris2021}). The survey will cover the sky at declination below +30$^{\circ}$ with an expected RMS noise level of 20-25~$\mu$Jy/beam. Unfortunately, the resolution of EMU is relatively poor ($\sim$12\arcsec) and the separation from core and extended emission may not be always possible. A follow-up of the detected sources could be necessary in order to estimate the correct flux density from the core.

An outstanding improvement, both in sensitivity and angular resolution, will be possible thanks to the wide-field continuum surveys that will be carried out by SKA1-Mid.
The continuum sensitivity (RMS in 1~hour) is expected to be 2~$\mu$Jy beam$^{-1}$ at 1.4 GHz with a resolution of 0.4$\arcsec$ (\citealt{Braun2019}). With this kind of imaging sensitivity and resolution, it will be possible to robustly test 
the presence of obscuration at large viewing angles.

In Fig.~\ref{fig_first} we show the 5$\times$rms flux limit of the surveys mentioned above, showing their actual capability of distinguishing between different scenarios. Reaching flux densities of a few tens of $\mu$Jy should allow us to test the presence of obscuration at angles up to $\sim$30-35$^\circ$, where the ``standard" obscuring torus should be present.

An important point that should be considered when moving towards such low radio flux densities while keeping the same optical limit is that, at a certain flux level, we  start sampling the population of QSO whose radio emission is not powered by a relativistic jet but could be related, for instance, to an intense star-formation. 
These are the sources usually classified as ``radio-quiet" (RQ) on the basis of a low radio-loudness (R$<$1). 
While in radio-loud (RL) QSO, with radio-loudness above 10,  the presence of a relativistic jet is well established, in RQ QSO the origin of the radio emission is still a matter of debate.
In the upper side of Fig.~\ref{fig_first} we report the values of radio-loudness computed using the radio flux density limit on the X-axis and the optical flux density at 4400\AA\ corresponding to mag$_{drop}$=21 and assuming z=4.5, $\alpha_R$=0.1\footnote{$\alpha_R$=0.1 is the average spectral slope measured for the sources in the sample between 144/150MHz and 1.4~GHz using data from LoTSS (\citealt{Shimwell2022}) and TGSS (\citealt{Intema2017}). These indices are reported in Tab.~\ref{tab:sample}} and $\alpha_O$=0.44 (\citealt{VandenBerk2001}). This scale gives an approximate indication of the typical values of R expected at each radio flux limit. For instance, among the 73 sources in the FIRST sample discussed here, the majority (65) have R$>$10 in the typical range of RL QSO, the remaining having R between 1 and 10. In the deeper ongoing/planned surveys, like ASKAP-EMU, instead, we should expect a significant fraction of sources with R between 1 and 10 or even below 1. 
The sample based on the very deep SKA1-Mid surveys will be almost completely made up by objects with a value of R in the RQ range ($<$1). This is interesting since the simple discovery of high-z blazars at high flux densities ($>$10~mJy) implies the existence of many misaligned jetted high-z QSO well in the ``RQ regime" (the actual number depending on the importance of the extended emission as explained above). Recent results by \citet{Sbarrato2021a} suggest that some high-z QSO with a radio-loudness well below the RL/RQ threshold do actually show evidence for the presence of a misaligned jet, putting into question the simple use of the radio-loudness parameter to discriminate between jetted and non-jetted QSO, at least in this range of redshift.

In any case, the overlap between misaligned jetted QSO and non-jetted QSO, whose radio emission could be due to star-formation or other mechanisms, is a potential issue since the possible inclusion of non-jetted QSO in the plot shown in Fig.~\ref{fig_first} is expected to produce a steepening of the curve below a certain flux limit thus affecting the final results. In principle, it should be possible to distinguish between the two classes of sources by studying the radio morphology (at arcsec and  sub-arcsecond scales) as demonstrated by \citet{Sbarrato2021a} or using the value of Brightness Temperature measured in high resolution radio images (see e.g. \citealt{Morabito2022}). 
In this context, wide-field VLBI surveys would be ideal to assess the presence of a jet (e.g. \citealt{Radcliffe2021a}). 
Also, the observed radio luminosity can be used to distinguish QSO whose radio emission is powered by a jet from those powered by star-formation. 
For example, all the sources in the FIRST sample with a radio-loudness parameter between 1 and 10 have radio powers at 1.4~GHz between 10$^{26}$ and 10$^{27}$ W Hz$^{-1}$  i.e. well within the typical range of power observed in jetted QSO, in spite of their relatively low radio-loudness values.

\section{Conclusions}
We have discussed and further developed a method, firstly proposed by \citet{Ghisellini2016}, to evaluate the presence of obscuration in jetted QSO in the early Universe (z$\geq$4), based on the number of blazars observed in a flux limited sample. We have applied this method to two well-defined samples of high-z type~1 jetted QSO containing, in total, 87 independent z$\geq$4 radio-emitting QSO, including 31 sources classified as blazars on the basis of the X-ray emission.  

The first sample (the CLASS high-z sample), containing 23 sources, is characterized by a high radio flux density limit (30~mJy at 5~GHz) while the second one is based on the combination of the last data-release of SDSS (DR17) and FIRST and contains 73 sources (9 sources are in common with the CLASS sample). This sample has a much deeper radio limit (0.5 mJy) compared to CLASS  but it has the same optical limit (mag$_{drop}$=21). 

The main results can be summarized as follows:

\begin{itemize}
    \item In the CLASS sample, blazars represent the large majority ($\sim$85\%) of the sources, with only a small fraction of misaligned objects. Our analysis shows that this dominance of oriented sources is consistent with the predictions of the beaming model and, therefore, it is likely due to the high radio-to-optical flux limit ratio of the CLASS sample that favours the selection of oriented sources, and it is not caused by the obscuration along large lines-of-sight; 

    \item Using the FIRST sample which contains a larger fraction of misaligned sources compared to CLASS we do not observe any significant departure from the beaming model predictions, although a possible deviation is hinted at low flux densities ($\sim$0.5~mJy). 
     It is possible, however, that this suggested trend is simply due to the incompleteness of the FIRST survey at such low flux density limit;

    \item Since a reliable blazar/non-blazar classification based on X-ray data is only present for the radio brightest sources in the FIRST samples (i.e. for $\sim$40\% of the objects) we cannot exclude that a discrepancy may appear once all the sources in the sample are correctly classified. However, considering that only sources with a relatively high radio-loudness are reasonable blazar candidates, the impact of the unclassified sources is quite marginal. Swift-XRT observations of the  objects in the sample with the highest radio-loudness are in progress. In addition, the first data release of the {\it eROSITA} All Sky Survey (\citealt{Merloni2012a}), expected in the next months, will certainly help to classify the sources since even a non-detection can be used to exclude their blazar nature;

    \item Next generation radio surveys, like ASKAP EMU and, eventually, those carried out with SKA1-Mid, will be able to sample much deeper radio limits and, therefore, much larger observing angles compared to the existing surveys. At some point, we do expect to observe a significant departure from the predictions, due to the presence of the "standard" obscuring torus. At these very low flux densities (tens of $\mu$Jy), however, distinguishing between misaligned QSO and  RQ QSO, i.e. sources not powered by a radio jet, could be challenging as the radio-loudness parameter will no longer be expected to be useful in separating the two types of sources.
     High resolution radio follow-up will be instrumental to distinguish the two classes of AGNs.
    Another critical point will be separating the core emission from that coming from isotropic extended structures, since at large observing angles the former will become less relevant. Again, using data with good resolution and at the highest possible frequency will be fundamental for a reliable analysis.
\end{itemize}

Assessing the fraction of optically absorbed jetted high-z QSO is also fundamental to establish how common powerful relativistic jets were in the early Universe. If the fraction of optically absorbed sources is similar to that estimated in high-z radio-quiet QSO ($\sim$70\%) the number of jetted QSO could be much higher compared to the local Universe, representing up to 90\% of the QSO population. Conversely, if jetted QSO represent only $\sim$10\% of the total population, as in the local Universe, the fraction of obscured jetted QSO could be much higher than expected, between 87\% and 97\%. Deep radio surveys will be instrumental to distinguish between these two possible (and intriguing) scenarios.

\begin{acknowledgements}
We acknowledge financial contribution from the agreement ASI-INAF n. I/037/12/0 and n. 2017-14-H.0 and from INAF under PRIN SKA/CTA FORECaST. We acknowledge financial support from INAF under the project ``QSO jets in the early Universe", Ricerca Fondamentale 2022 and under the project ``Testing the obscuration in the Early Universe", Ricerca Fondamentale 2023. CS acknowledges financial support from INAF (Grants 1.05.12.04.04). This work has been partially supported by the ASI-INAF program I/004/11/4.

This work is based on FIRST and SDSS data.
Funding for the Sloan Digital Sky Survey V has been provided by the Alfred P. Sloan Foundation, the Heising-Simons Foundation, the National Science Foundation, and the Participating Institutions. SDSS acknowledges support and resources from the Center for High-Performance Computing at the University of Utah. The SDSS web site is \url{www.sdss.org}.
\end{acknowledgements}


%\begin{thebibliography}{}
\bibliographystyle{aa}
\bibliography{aanda.bbl}

%\end{thebibliography}
\appendix
\section{Blazars and non-blazars in a flux limited sample}
We tested the method presented by \citet{Ghisellini2019} and discussed in Sect.~2 (equation~1) using numerical simulations. Equation~1 gives the total number of jetted QSO expected in a flux-limited sample given the flux densities of all the blazars discovered in the same sample. 
To test this formula, we started from a population of un-beamed sources with
an intrinsic (i.e. rest-frame) flux density ranging from 0.01 mJy to 100 mJy\footnote{we have used different flux distributions, i.e. we started with luminosity functions with different slopes, but we found that the result does not depend on the assumed function.} and randomly associated to each source a viewing angle. We then applied the beaming factor to these sources
assuming different values of $\Gamma$ (5, 10 and 15) and p (2 and 3) and considered  only the sources above a certain flux limit (we use here 30~mJy that is the limit of the CLASS sample, but the results can be re-scaled to whichever value). We have then classified as blazars all the sources with $\theta<1/\Gamma$ and applied equation~1 to recover the expected number of sources, both blazars and misaligned objects, above the flux limit. This value is then compared to the actual number of objects above the same limit.

We note that in these simulations we have assumed that there is no extended, un-beamed
emission and, therefore, we expect that the value of N$_{tot}$ provided by equation~1 should match exactly the actual value of sources given as input.
In real data, extended emission may give a contribution although, as discussed in Sect.~2, we do not expect that this is relevant in the samples studied in this work.

\begin{figure}
   \centering
    \includegraphics[width=8cm, angle=0]{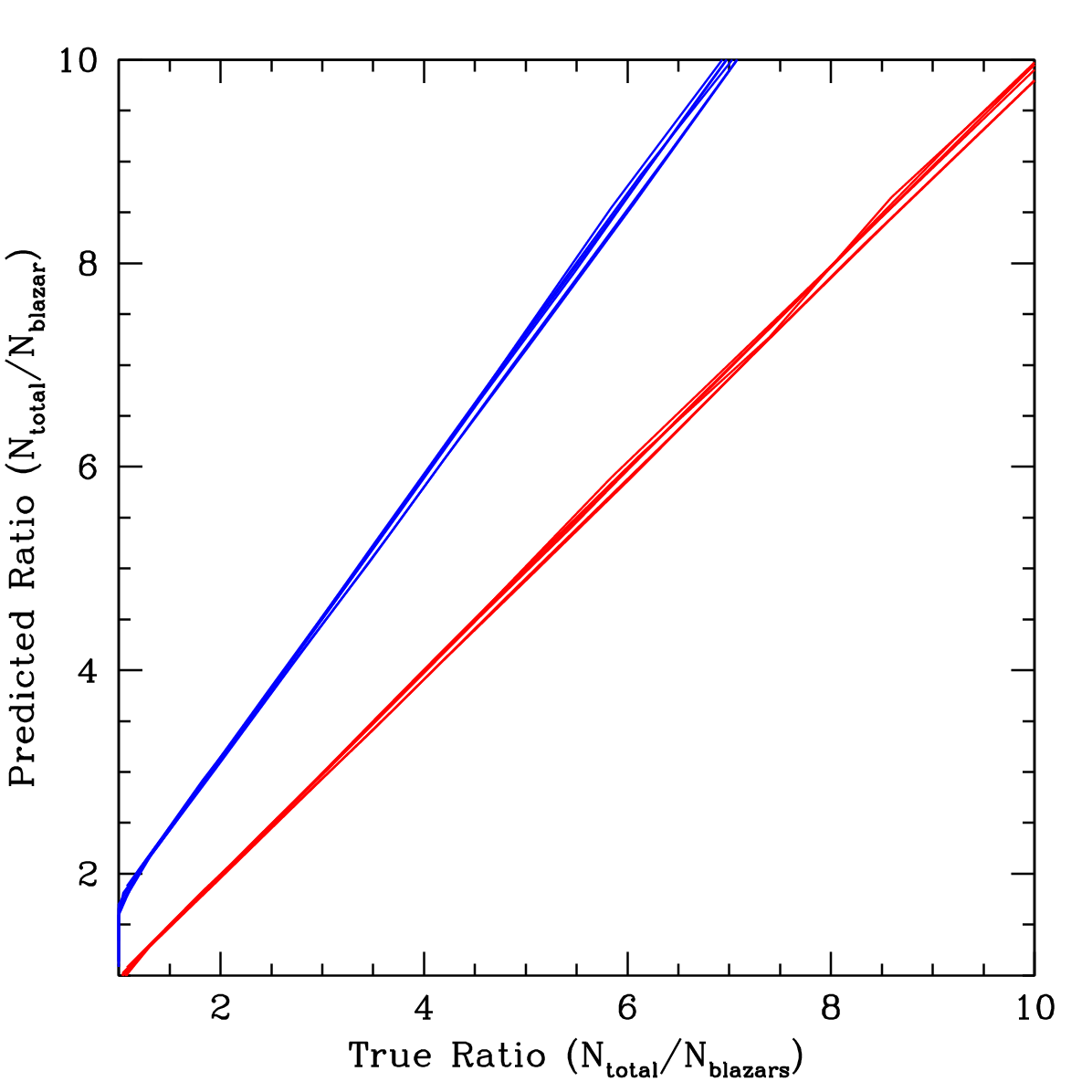}
   \caption{Results of a simulation where N jets of equal radio flux density 
have been randomly oriented in the sky and relativistically beamed according to the 
angle ($\theta$) between the jet and the observer. We assume different values of the beaming
parameters, i.e. $\Gamma$=5, 10, 15 and p=2 and 3 and different starting values of the un-beamed radio flux density (from 0.01 to 100 mJy). We then imposed a flux limit of 30~mJy and 
classified all the sources with $\theta<1/\Gamma$ 
as blazars. These combinations of values yield different values (from 1 to 10) 
of the ratio between total number of sources
(blazar+misaligned objects) and 
blazars that are included in the simulated sample. 
We finally applied Eq~1 and Eq~2 to the 
simulated blazar sample (assuming the correct value of p)  to 
recover the expected total number of sources above the flux limit. In the figure, we compare the predicted total/blazar number ratio with the true one.  
Equation~1 (blue line) systematically overestimates the result by a factor $\sim$1.4 
while equation~2 (red line) provides the correct results
}
              \label{test_gg_all2}
    \end{figure}

Depending on the assumed values of
the intrinsic flux density, $\Gamma$ and p we obtain different values of total-to-blazar number ratio. 
Fig.~\ref{test_gg_all2} shows the results of the comparison between the predicted ratio derived from
equation~1 and the real one. It appears that equation~1 systematically overestimates 
(by a factor $\sim$1.4) the true total number  of sources in the sample (blue lines). 
The overestimate is nearly independent of the assumed value of the input parameters. 

The origin of this discrepancy is likely related to the fact that  equation~1 was derived assuming that all the blazars in the sample 
are oriented exactly at an angle equal to 1/$\Gamma$. At this angle, the Doppler factor\footnote{The Doppler factor is defined as $\delta=[\Gamma(1-\beta cos\theta)]^{-1}$, where 
$\Gamma$ and $\beta$ are the Lorentz factor and the ratio between the bulk velocity and the 
light speed respectively} is equal to $\Gamma$. This is clearly an approximation,  
since the actual viewing angles of a sample of blazars are expected to be distributed
between 0$^\circ$ and 1/$\Gamma$. A more correct number should be given by the average value of 
Doppler factor computed on all the sources observed within 1/$\Gamma$. 
This average value is expected to be larger than $\Gamma$. 
The average Doppler factor for a sample of blazars is given by:

\begin{equation}
\frac{\int_{0}^{1/\Gamma} \delta(\theta) sin(\theta) d\theta}{1-cos(1/\Gamma)}\sim1.385\Gamma
\end{equation}

Including this correction in the derivation of equation~1 presented in \citet{Ghisellini2016}, 
we obtain:

\begin{equation}
N_{tot} \sim \sum_{i=1}^N {[1.44(\frac{S_i}{S_{lim}})^{1/p} -1]}
\end{equation}
%______________________________________________ Gamma_1 (lg rho, lg e)
%-------------------------------------------------
Again, we tested this relation through our numerical simulations (red line in Fig~\ref{test_gg_all2}) and we
found that it correctly recovers the true total number of sources.
%-----------------------------------------------
\section{Reliability of the classifications}
As mentioned in Sect.~4.1, the classification of a source as blazar is not straightforward, and it may depend on the adopted method. Recently, \citet{Krezinger2022a} 
pointed out some discrepancies between the classification based on the SED analysis and the one based on VLBI data.
Among the high-z QSO in the samples considered here, there are 16 objects with a classification from the literature based on VLBI data. While in most (70\%) cases the two classifications agree, in 5 cases there seems to be a disagreement. These cases are discussed below.

{\bf J083946.2+511202} - This source is classified as blazar based on the $\tilde{\alpha}_{ox}$ (=1.25). The blazar nature is also suggested by the analysis of the SED (\citealt{Sbarrato2013a}). Quite consistently to this classification, VLBI data (\citealt{Cao2017}) show a flat-spectrum, variable and quite compact core. However, the T$_b$ estimated from the deconvolved size (3.0$\pm0.5\times$10$^{10}$ K) is close but below the equipartition limit. This is likely a borderline source (also from the $\tilde{\alpha}_{ox}$  point of view) whose exact orientation could be near the one expected for blazars.

{\bf J103717.7+182303} - The value of $\tilde{\alpha}_{ox}$ of this object is similar to the one computed for the previous source (1.28) and is suggestive of a blazar nature. From the analysis of the SED \citet{Sbarrato2022} conclude that this source is either a blazar or a source viewed at the border of the definition.
Again, VLBI data seem to resolve the emission finding a relatively low value of T$_b$ (2.6$\pm$0.5$\times$10$^9$ K), below the equipartition limit (\citealt{Krezinger2022a}). As for J083946.2+511202, this could be a source at the border of the blazar definition.

{\bf J114657.8+403708} - Also this object is classified as blazar on the basis of the X-ray data ($\tilde{\alpha}_{ox}$=1.31) and of the SED (\citealt{Ghisellini2014}) but the value of T$_b$ derived from VLBI data (4.5$\pm$0.3$\times$10$^9$ K) is below the equipartition limit (\citealt{Frey2010}).

{\bf J123142.1+381659} - 
The value of  $\tilde{\alpha}_{ox}$
computed for this object (1.18) indicates a blazar nature. The same classification is reported in \citet{Sbarrato2022} while \citet{Krezinger2022a} suggest that the source is misaligned on the basis of the relatively low T$_b$ (1.36$\pm$0.27$\times$10$^9$ K).

{\bf J142048.0+120546} - Again, we classified this source as blazar due to its low $\tilde{\alpha}_{ox}$ (1.21) and this classification is supported by the analysis of the SED (\citealt{Sbarrato2014a}). VLBI observations with EVN, instead, reveal a double morphology on a scale of 1.33$\arcsec$, which is too compact to be detected by FIRST data (\citealt{Cao2017}). The most compact and brightest component, that is positionally coincident with the optical position of the QSO, has a relatively high T$_b$ (4.0$\pm$1.0$\times$10$^9$ K) but below the equipartition limit. 

To summarize, all the 5 cases of inconsistent classification between the two methods are quite similar and include sources with values of $\tilde{\alpha}_{ox}$
below but close to the limit of 1.355 (typically 1.2-1.3) and show brightness temperature relatively high (above 10$^9$K) but below the equipartition limit (0.3-5$\times$10$^{10}$ K). At the same time, we have other 4 sources with similar values of $\tilde{\alpha}_{ox}$ ($\sim$1.2-1.3) but for which VLBI observations confirm the presence of boosting. 

All the remaining 7 objects with VLBI data, i.e. sources with either a very low value of $\tilde{\alpha}_{ox}$ (between 0.8 and 1.1, 4 objects) or a value above 1.355 (classified as non-blazars, 3 objects) have a consistent classification from VLBI data.

We conclude that the two methods are in  reasonably good agreement, except for sources that are borderline ($\tilde{\alpha}_{ox}$ between 1.2 and 1.3). In this range, $\sim$50\% of the sources have a VLBI classification that seems to contradict the one inferred from the X-ray data. As already said, these are cases that require a more careful analysis that takes into account all the possible pieces of information available. We stress here that also the VLBI classification can be uncertain when the measured extensions, critical to determine the value of T$_b$, are only fractions of the synthesized beam and when the flux densities are relatively low compared to the local RMS. Only a synergic use of all the available multiwavelength data can provide the best possible classification.

In any case, as discussed in Sect.~4.2, if we do not consider as blazars the sources with controversial classification, the results presented in this paper do not change significantly. 

\section{The samples}
In Tab~\ref{tab:sample} we present the complete list of sources used in this paper. For each source in the CLASS and FIRST samples, we give the following quantities. Column~1: name based on the optical coordinates. Column~2-3: optical position. Columns~4: redshift derived from \citet{Caccianiga2019} or from SDSS except for four objects: $^1$\citet{Banados2015a}, $^2$\citet{Gloudemans2022}, $^3$\citet{Banados2021}, $^4$\citet{Stern2000}. Column~5: mag$_{drop}$ corrected for Galactic reddening. The filter is specified between parenthesis. Magnitudes are taken from Pan-STARRS1 survey (PS1, \citealt{Chambers2016}) or from the DECam Local Volume Exploration Survey (DELVE, \citealt{Drlica-Wagner2021, Drlica-Wagner2022}); Column~6: 1.4~GHz peak flux density from FIRST. Column~7: 5~GHz flux density from GB6 catalogue (\citealt{Gregory1996}). Column~8: non-simultaneous radio spectral index (S$_{\nu}\propto\nu^{-\alpha}$) from 1.4~GHz and 144/150~MHz using FIRST and LoTSS (\citealt{Shimwell2022}) or TGSS (\citealt{Intema2017}) integrated flux densities. Column~9: sample: F=FIRST sample, C=CLASS sample. Column~10: Log of radio-loudness computed using 1.4~GHz flux density, and assuming the radio slope given in column~8 or assuming $\alpha_r$=0.1, and 
using the near-infrared magnitude at 3.4~micron (W1) from Wide-field Infrared Survey Explorer (NEOWISE, \citealt{Mainzer2014}) which provides the photometric point closest to 4400\AA\ rest-frame wavelength for z$\geq$4 sources. To obtain the monochromatic flux at 4400\AA\ we use the spectral index computed between W1 and z-filter (or y-filter for z$>$6.3 sources). If W1 is not available, we compute R using the z magnitude and assuming $\alpha_0$=0.44 (\citealt{VandenBerk2001});
Column~11: The two-points-spectral index from radio to X-rays ($\tilde{\alpha}_{ox}$). This is either taken from \citet{Ighina2019a}, for the sources in the CLASS sample, or computed using the available X-ray fluxes and the magnitudes in z-filter. As for the radio-loudness parameter, we use the optical spectral index computed between W1 and z-filter, if available, or assume $\alpha_0$=0.44. The error on this parameter is mostly due to the uncertainty on the optical/UV slope and on the X-ray fluxes. Considering the range of values of slopes reported in Table~5 of  \citet{VandenBerk2001} (0.33-0.93) and the typical errors on the X-ray fluxes, we derive a typical error on $\tilde{\alpha}_{ox}$ of -0.03, +0.06.  Column~12: classification based on X-ray data: b=blazar ($\tilde{\alpha}_{ox} \leq$1.355), nb=non-blazar ($\tilde{\alpha}_{ox}>$1.355). 
\footnotesize

%%% START TABLE
\begin{table*}
\scriptsize 
 \centering
 \caption{The sample of high-z jetted QSO}
 \label{tab:sample}
 \begin{tabular}{lrrllrrrccrl}
  \hline
name                 &  RA        &   DEC      &  z     &  mag$_{drop}$     & $S_{FIRST}^{pk}$ & $S_{GB6}$ & $\alpha_R$ & sample & logR & $\tilde{\alpha}_{ox}$ & class \\
                     &  (J2000)   &   (J2000)  &        &                   &  [mJy]  &  [mJy]   &     &        &       &       \\
         (1)         &  (2)       &   (3)      &  (4)   &    (5)            &  (6)    &  (7)     & (8) &  (9)   & (10)  & (11)  \\
  \hline
J001115.2+144601 &   2.81348 &  14.76718 &  4.97  & 18.20$\pm$0.01 ( i          ) &   24.0 &     31 &          0.14 & C &  2.1 & 1.17 & b\\
J003126.8+150739 &   7.86167 &  15.12764 &  4.28  & 20.00$\pm$0.01 ( r          ) &   43.0 &     93 & - & C &  2.8 & 1.40 & nb\\
J012126.1+034706 &  20.35898 &   3.78515 &  4.13  & 18.68$\pm$0.01 ( r          ) &   73.0 &     51 & - & C &  2.4 & 1.41 & nb\\
J012201.9+031002 &  20.50792 &   3.16733 &  4.00  & 20.85$\pm$0.02 ( r          ) &  106.1 &     96 &         -0.07 & C &  3.6 & 0.79 & b\\
J083549.4+182520 & 128.95594 &  18.42225 &  4.41  & 20.79$\pm$0.02 ( r          ) &   51.5 &     40 &          0.21 & C &  3.2 & 1.03 & b\\
J083946.2+511202 & 129.94259 &  51.20080 &  4.40  & 19.35$\pm$0.01 ( r          ) &   40.5 &     51 &         -0.13 & C &  2.2 & 1.25 & b\\
J090132.6+161506 & 135.38604 &  16.25190 &  5.63$^1$ & 20.69$\pm$0.03 ( z          ) &    3.0 & <30 & - & F &  1.9 & 1.33 & b\\
J090630.7+693030 & 136.62813 &  69.50856 &  5.47  & 19.77$\pm$0.03 ( z          ) & - &    106 & - & C &  3.2 & 1.05 & b\\
J091316.5+591921 & 138.31899 &  59.32267 &  5.12  & 20.58$\pm$0.02 ( i          ) &   17.4 & <30 &         -0.33 & F &  3.2 & 1.60 & nb\\
J091824.3+063653 & 139.60158 &   6.61481 &  4.19  & 19.60$\pm$0.01 ( r          ) &   25.9 &     36 & - & CF &  2.1 & 1.26 & b\\
J092132.7+185654 & 140.38648 &  18.94860 &  4.52  & 21.00$\pm$0.04 ( i          ) &    4.3 & <30 & - & F &  2.0 & - & -\\
J092709.6+282229 & 141.79038 &  28.37476 &  4.30  & 20.19$\pm$0.01 ( r          ) &    1.9 & <30 &         -0.09 & F &  1.7 & - & -\\
J094004.8+052630 & 145.02000 &   5.44194 &  4.50  & 20.75$\pm$0.03 ( i          ) &   55.7 & <30 &          0.87 & F &  3.4 & 1.09 & b\\
J094409.5+100656 & 146.03968 &  10.11575 &  4.76  & 19.17$\pm$0.01 ( i          ) &    0.6 & <30 & - & F &  0.7 & - & -\\
J100303.4+112209 & 150.76435 &  11.36938 &  4.58  & 20.83$\pm$0.02 ( i          ) &    2.3 & <30 & - & F &  2.2 & - & -\\
J100645.5+462717 & 151.68999 &  46.45479 &  4.44  & 20.61$\pm$0.02 ( r          ) &    6.0 & <30 &         -0.39 & F &  2.2 & - & -\\
J100903.3+174950 & 152.26404 &  17.83067 &  4.05  & 20.77$\pm$0.02 ( r          ) &    0.5 & <30 & - & F &  1.5 & - & -\\
J101337.8+351849 & 153.40779 &  35.31384 &  5.03$^2$ & 20.58$\pm$0.02 ( i          ) &    1.6 & <30 &          0.20 & F &  1.5 & - & -\\
J101506.6+281757 & 153.77790 &  28.29942 &  4.15  & 19.85$\pm$0.01 ( r          ) &    1.1 & <30 &          0.12 & F &  1.3 & - & -\\
J102043.8+000105 & 155.18258 &   0.01830 &  4.19  & 20.40$\pm$0.01 ( r          ) &    1.7 & <30 & - & F &  1.4 & - & -\\
J102249.2+130125 & 155.70530 &  13.02363 &  4.03  & 19.36$\pm$0.01 ( r          ) &    2.6 & <30 & - & F &  0.7 & - & -\\
J102343.1+553132 & 155.92974 &  55.52566 &  4.46  & 20.59$\pm$0.02 ( r          ) &    2.0 & <30 &         -0.17 & F &  1.0 & - & -\\
J102623.6+254259 & 156.59845 &  25.71651 &  5.25  & 19.92$\pm$0.01 ( i          ) &  230.8 &    142 &          0.25 & CF &  3.6 & 1.15 & b\\
J103418.6+203300 & 158.57771 &  20.55006 &  5.01  & 19.75$\pm$0.01 ( i          ) &    4.0 & <30 & - & F &  1.5 & - & -\\
J103446.5+110214 & 158.69393 &  11.03736 &  4.27  & 18.82$\pm$0.01 ( r          ) &    0.9 & <30 & - & F &  0.8 & - & -\\
J103601.0+500831 & 159.00429 &  50.14215 &  4.48  & 20.08$\pm$0.01 ( r          ) &    9.2 & <30 &          0.35 & F &  1.7 & - & -\\
J103717.7+182303 & 159.32386 &  18.38418 &  4.04  & 19.89$\pm$0.01 ( r          ) &   13.6 & <30 & - & F &  2.3 & 1.28 & b\\
J104007.3+161809 & 160.03066 &  16.30265 &  4.08  & 20.96$\pm$0.02 ( r          ) &    3.7 & <30 & - & F &  1.4 & - & -\\
J104742.5+094744 & 161.92741 &   9.79581 &  4.25  & 20.51$\pm$0.01 ( r          ) &   18.9 & <30 &          0.17 & F &  2.4 & - & -\\
J105756.2+455553 & 164.48446 &  45.93140 &  4.13  & 17.63$\pm$0.01 ( r          ) &    1.1 & <30 &          1.14 & F &  0.2 & 1.56 & nb\\
J110201.9+533912 & 165.50797 &  53.65351 &  4.31  & 20.35$\pm$0.01 ( r          ) &    4.5 & <30 &          0.47 & F &  2.1 & - & -\\
J110549.9+290225 & 166.45798 &  29.04030 &  4.07  & 20.49$\pm$0.02 ( r          ) &    3.9 & <30 &          0.40 & F &  2.2 & - & -\\
J111639.7+092612 & 169.16552 &   9.43668 &  4.13  & 20.33$\pm$0.01 ( r          ) &    1.4 & <30 & - & F &  1.6 & - & -\\
J112925.3+184624 & 172.35560 &  18.77340 &  6.82$^3$ & 20.74$\pm$0.09 ( y          ) &    1.0 & <30 & - & F &  1.8 & - & -\\
J113729.4+375224 & 174.37262 &  37.87340 &  4.24  & 19.95$\pm$0.01 ( r          ) &    2.7 & <30 &         -0.09 & F &  1.8 & - & -\\
J114657.8+403708 & 176.74084 &  40.61911 &  5.01  & 19.40$\pm$0.01 ( i          ) &   12.4 & <30 &         -0.67 & F &  1.9 & 1.31 & b\\
J121134.3+322615 & 182.89330 &  32.43757 &  4.14  & 19.56$\pm$0.01 ( r          ) &    3.7 & <30 &          0.44 & F &  1.2 & - & -\\
J121329.0+181029 & 183.37114 &  18.17476 &  4.50  & 20.08$\pm$0.01 ( i          ) &    2.3 & <30 & - & F &  2.1 & - & -\\
J123142.1+381659 & 187.92554 &  38.28311 &  4.14  & 20.11$\pm$0.02 ( r          ) &   20.4 & <30 &          0.36 & F &  2.7 & 1.18 & b\\
J123237.4+520343 & 188.15623 &  52.06218 &  4.61  & 20.55$\pm$0.03 ( i          ) &    3.6 & <30 &         -0.42 & F &  2.2 & - & -\\
J123239.2+525250 & 188.16373 &  52.88083 &  4.34  & 18.44$\pm$0.01 ( r          ) &    0.5 & <30 & - & F &  0.3 & - & -\\
J123604.1+030341 & 189.01747 &   3.06163 &  4.58  & 19.98$\pm$0.01 ( i          ) &    2.0 & <30 & - & F &  1.5 & - & -\\
J124230.5+542257 & 190.62744 &  54.38261 &  4.73  & 19.82$\pm$0.01 ( i          ) &   19.7 & <30 &          0.16 & F &  2.4 & 1.41 & nb\\
J124943.6+152707 & 192.43196 &  15.45196 &  4.02  & 19.34$\pm$0.01 ( r          ) &    1.8 & <30 & - & F &  1.2 & - & -\\
J130002.1+011823 & 195.00902 &   1.30641 &  4.61  & 18.88$\pm$0.01 ( i          ) &    2.7 & <30 & - & F &  1.1 & 1.27 & b\\
J130738.8+150752 & 196.91180 &  15.13114 &  4.08  & 19.98$\pm$0.01 ( r          ) &    3.4 & <30 & - & F &  1.8 & - & -\\
J130940.7+573309 & 197.41960 &  57.55277 &  4.28  & 19.57$\pm$0.01 ( r          ) &   10.7 & <30 &         -0.15 & F &  1.6 & 1.35 & b\\
J131121.3+222738 & 197.83886 &  22.46075 &  4.61  & 20.45$\pm$0.02 ( i          ) &    6.5 & <30 & - & F &  2.1 & 1.37 & nb\\
J131814.0+341805 & 199.55846 &  34.30156 &  4.88  & 19.13$\pm$0.01 ( i          ) &    3.7 & <30 &         -0.49 & F &  1.0 & - & -\\
J132512.4+112329 & 201.30206 &  11.39160 &  4.41  & 19.45$\pm$0.01 ( r          ) &   69.4 &     62 & - & CF &  2.6 & 1.31 & b\\
J133422.6+475033 & 203.59431 &  47.84267 &  4.95  & 19.59$\pm$0.01 ( i          ) &    0.6 & <30 &          0.98 & F &  1.1 & 1.48 & nb\\
J134811.2+193523 & 207.04690 &  19.58990 &  4.40  & 20.66$\pm$0.02 ( r          ) &   49.3 &     38 &          0.00 & CF &  3.0 & 1.23 & b\\
J135135.7+284014 & 207.89879 &  28.67078 &  4.72  & 19.66$\pm$0.01 ( i          ) &    3.2 & <30 &         -0.47 & F &  1.7 & - & -\\
J135554.5+450421 & 208.97733 &  45.07252 &  4.10  & 19.55$\pm$0.01 ( r          ) &    1.5 & <30 &         -0.58 & F &  1.5 & - & -\\
J140025.4+314910 & 210.10587 &  31.81961 &  4.64  & 20.20$\pm$0.01 ( i          ) &   20.2 & <30 &         -0.06 & F &  2.7 & 1.23 & b\\
J140034.0+173031 & 210.14186 &  17.50872 &  4.29  & 20.74$\pm$0.02 ( r          ) &    1.6 & <30 & - & F &  1.4 & - & -\\
J140850.9+020522 & 212.21215 &   2.08964 &  4.01  & 18.98$\pm$0.01 ( r          ) &    1.5 & <30 & - & F &  1.2 & 1.40 & nb\\
J141209.9+062406 & 213.04155 &   6.40191 &  4.36  & 20.00$\pm$0.01 ( r          ) &   43.0 &     34 &          0.43 & CF &  3.1 & 1.29 & b\\
J142048.0+120546 & 215.20004 &  12.09611 &  4.03  & 19.80$\pm$0.01 ( r          ) &   83.8 &     47 &          0.66 & CF &  3.2 & 1.21 & b\\
J142308.2+224158 & 215.78435 &  22.69946 &  4.32  & 19.83$\pm$0.01 ( r          ) &   35.1 &     22 & - & F &  2.7 & 1.33 & b\\
J142634.8+543622 & 216.64524 &  54.60634 &  4.85  & 19.90$\pm$0.01 ( i          ) &    4.3 & <30 &         -0.18 & F &  1.8 & - & -\\
J143023.7+420436 & 217.59891 &  42.07681 &  4.71  & 19.78$\pm$0.01 ( i          ) &  211.3 &    337 &         -0.10 & CF &  3.4 & 0.82 & b\\
J143413.0+162852 & 218.55441 &  16.48131 &  4.21  & 19.83$\pm$0.01 ( r          ) &    4.8 & <30 & - & F &  1.7 & - & -\\
J143942.9+012741 & 219.92911 &   1.46161 &  4.16  & 20.43$\pm$0.02 ( r          ) &    3.7 & <30 & - & F &  1.7 & - & -\\
J144231.7+011055 & 220.63216 &   1.18203 &  4.49  & 20.14$\pm$0.01 ( i          ) &    1.1 & <30 & - & F &  1.9 & 1.27 & b\\
J144720.5+164018 & 221.83575 &  16.67167 &  4.07  & 20.43$\pm$0.02 ( r          ) &    1.1 & <30 & - & F &  1.4 & - & -\\
J145224.2+335424 & 223.10106 &  33.90684 &  4.12$^4$ & 20.40$\pm$0.01 ( r          ) &    6.8 & <30 &         -0.22 & F &  1.8 & - & -\\
J145628.9+200727 & 224.12072 &  20.12419 &  4.26  & 19.51$\pm$0.01 ( r          ) &    0.7 & <30 & - & F &  0.6 & - & -\\
J145924.0+035622 & 224.85030 &   3.93951 &  4.13  & 20.54$\pm$0.01 ( r          ) &    3.3 & <30 & - & F &  1.6 & - & -\\
  \hline
 \end{tabular}
\end{table*}
\setcounter{table}{0}
\begin{table*}
\scriptsize 
 \centering
 \caption{continued}
 \label{tab:sample}
 \begin{tabular}{lrrllrrrccrl}
  \hline
  \hline
J150149.0+592252 & 225.45436 &  59.38119 &  4.65  & 18.84$\pm$0.01 ( i          ) &    4.6 & <30 &          0.71 & F &  1.5 & - & -\\
J150544.6+433824 & 226.43583 &  43.64019 &  4.68  & 18.82$\pm$0.01 ( i          ) &    5.0 & <30 &          0.11 & F &  1.2 & - & -\\
J150912.2+202653 & 227.30100 &  20.44832 &  4.06  & 20.85$\pm$0.02 ( r          ) &    6.3 & <30 & - & F &  2.4 & - & -\\
J151002.9+570243 & 227.51219 &  57.04539 &  4.31  & 20.48$\pm$0.03 ( r          ) &  248.1 &    292 &          0.11 & CF &  3.5 & 0.94 & b\\
J152028.1+183556 & 230.11726 &  18.59894 &  4.12  & 19.60$\pm$0.01 ( r          ) &    6.3 & <30 & - & F &  2.3 & $>$1.50 & nb\\
J152028.1+210039 & 230.11734 &  21.01101 &  4.55  & 20.12$\pm$0.02 ( i          ) &    1.1 & <30 & - & F &  1.9 & - & -\\
J152404.2+134417 & 231.01763 &  13.73821 &  4.79  & 19.32$\pm$0.01 ( i          ) &    0.5 & <30 & - & F &  0.7 & - & -\\
J152759.0+344118 & 231.99600 &  34.68840 &  4.32  & 20.92$\pm$0.03 ( r          ) &    4.7 & <30 &          0.30 & F &  2.4 & - & -\\
J153533.8+025423 & 233.89120 &   2.90650 &  4.39  & 20.43$\pm$0.01 ( r          ) &   78.8 &     53 &          0.40 & CF &  3.6 & 0.93 & b\\
J153830.7+424405 & 234.62799 &  42.73489 &  4.10  & 20.85$\pm$0.03 ( r          ) &   11.7 & <30 &         -0.22 & F &  2.6 & $>$1.35 & nb\\
J154824.0+333500 & 237.10006 &  33.58336 &  4.68  & 20.77$\pm$0.02 ( i          ) &   37.6 & <30 &          0.54 & F &  3.6 & 0.96 & b\\
J161216.7+470253 & 243.06981 &  47.04823 &  4.35  & 20.62$\pm$0.02 ( r          ) &   52.3 &     30 &          0.18 & C &  2.9 & 1.39 & nb\\
J162957.2+100023 & 247.48867 &  10.00653 &  5.00  & 20.45$\pm$0.02 ( i          ) &   51.5 &     33 &          0.50 & C &  3.5 & 1.09 & b\\
J164854.5+460327 & 252.22721 &  46.05761 &  5.36  & 20.29$\pm$0.01 ( i          ) &   32.0 &     30 &          0.21 & C &  2.8 & 1.34 & b\\
J171105.5+383004 & 257.77308 &  38.50121 &  4.00  & 20.41$\pm$0.02 ( r          ) &   49.1 &     36 &         -0.22 & C &  3.0 & 1.15 & b\\
J222032.5+002537 & 335.13542 &   0.42708 &  4.20  & 20.04$\pm$0.01 ( r          ) &   61.9 &     37 &          0.91 & C &  3.3 & 1.26 & b\\
J231448.7+020151 & 348.70296 &   2.03086 &  4.11  & 19.67$\pm$0.01 ( r          ) &  117.8 &     84 &         -0.05 & C &  3.1 & 1.25 & b\\
J235758.5+140201 & 359.49396 &  14.03384 &  4.33  & 20.25$\pm$0.01 ( r          ) & - &     78 &          0.18 & C &  3.3 & 1.02 & b\\
  \hline
 \end{tabular}
\end{table*}
%%% END TABLE

\end{document}